\documentclass[prc,twocolumn,showpacs,showkeys]{revtex4}
\usepackage{epsf}
\usepackage{amssymb}
\usepackage{dcolumn}
\def\g{\gamma}
\def\m{\mu^2}
\def\G{\Gamma}

\def\am{(\alpha_1-\beta_1)}
\def\ap{(\alpha_1+\beta_1)}
\def\amn{\am_{\pi^0}}
\def\amc{\am_{\pi^{\pm}}}
\def\apc{\ap_{\pi^{\pm}}}
\def\apn{\ap_{\pi^0}}
\def\bm{(\alpha_2-\beta_2)}
\def\bp{(\alpha_2+\beta_2)}
\def\bmn{\bm_{\pi^0}}
\def\bmc{\bm_{\pi^{\pm}}}
\def\bpc{\bp_{\pi^{\pm}}}
\def\bpn{\bp_{\pi^0}}
\def\gg{\g \g\to \pi^0 \pi^0}
\def\s{\sigma}
\def\ggc{\g\g\to\pi^+\pi^-}
\def\gp{\g p\to\g\pi^+n}

\def\sigc{\frac{d\s_{\ggc}}{d\Omega}}
\def\tg{\theta^{*}}

\def\mpp{M_{++}}
\def\mm{M_{+-}}
\def\sp{s'}
\def\tp{t'}
\def\ov{\overline}
\def\unit{10^{-4} {\rm fm}^3}
\def\unitq{10^{-4} {\rm fm}^5}
\def\be{\begin{equation}}
\def\ee{\end{equation}}
\def\beq{\begin{eqnarray}}
\def\eeq{\end{eqnarray}}
\begin{document}
\title{Determination of $\pi^{\pm}$ meson polarizabilities from
the $\ggc$  process}
\vskip 0.5cm

\author{L.V.~Fil'kov}
\email[E-mail: ]{filkov@sci.lebedev.ru}
\author{V.L.~Kashevarov}
\email[E-mail: ]{kashev@kph.uni-mainz.de}
\affiliation{Lebedev Physical Institute, Leninsky Prospect 53,
Moscow 119991, Russia}
\vskip 1cm

\begin{abstract}
A fit of the experimental data to the total cross section of the
process $\ggc$ in the energy region from threshold to 2500 MeV
has been carried out using dispersion relations with subtractions for the
invariant amplitudes, where the dipole and the quadrupole polarizabilities
of the charged pion are free parameters. As a result, the sum and the
difference of the electric and magnetic dipole and  quadrupole
polarizabilities of the charged pion have been found:
\mbox{$\apc=(0.18^{+0.11}_{-0.02})\times\unit$},
\mbox{$\amc=(13.0^{+2.6}_{-1.9})\times\unit$},
\mbox{$\bpc=(0.133\pm 0.015)\times\unitq$},
\mbox{$\bmc=(25.0^{+0.8}_{-0.3})\times\unitq$}.
These values agree with the dispersion sum rule predictions.
The value found for the difference of the dipole polarizabilities is consistent
with the results obtained from scattering of high energy $\pi^-$ mesons off
the Coulomb field of heavy nuclei [Yu.M. Antipov {\em et al.}, Phys. Lett.
{\bf B121}, 445 (1983)] and from radiative
$\pi^+$ photoproduction from the proton at MAMI [J. Ahrens {\em et al.}, 
Eur. Phys. J. A {\bf 23}, 113 (2005)], whereas it is at
variance with the recent calculations in the framework of chiral perturbation
theory.
\end{abstract}
\vskip 0.3cm


\maketitle

\section{Introduction}

Pion polarizabilities are fundamental structure parameters characterizing
the behavior of the pion in an external electromagnetic field.
The dipole and quadrupole polarizabilities arise as ${\cal O}(\omega^2)$
and ${\cal O}(\omega^4)$ terms, respectively, in the expansion of the
non-Born amplitude of Compton scattering in powers of the initial photon energy
$\omega$. In terms of the electric $\alpha_l$ ($l=$1, 2) and magnetic
$\beta_l$ dipole and quadrupole polarizabilities, the corresponding effective
interactions of ${\cal O}(\omega^2)$ and
${\cal O}(\omega^4)$ have the forms \cite{lvov,holst1}:
\be
H^{(2)}_{eff}=-\frac12\,4\pi\,(\alpha_1\,\vec{E}^2+\beta_1\,\vec{H}^2),
\ee
\be
H^{(4)}_{eff}=-\frac{1}{12}4\pi\left(\alpha_2\,E^2_{ij}+
\beta_2\,H^2_{ij}\right),
\ee
where
\be
E_{ij}=\frac12\,\left(\nabla_i\,E_j+\nabla_j\,E_i\right), \;
H_{ij}=\frac12\,\left(\nabla_i\,H_j+\nabla_j\,H_i\right)
\ee
are the quadrupole strengths of the electric and magnetic fields.

The dipole polarizabilities ($\alpha_1$ and $\beta_1$) of the pion
measure the response of the pion to quasistatic electric and magnetic
fields. In contrast, the parameters $\alpha_2$ and $\beta_2$
measure the electric and magnetic quadrupole moments induced in the pion
in the presence of an applied field gradient.
In what follows, the dipole and quadrupole polarizabilities are given in
units $\unit$ and $\unitq$, respectively.

The values of the pion polarizabilities are very sensitive to predictions of
different theoretical models. Therefore, an accurate experimental determination
of these parameters is very important for testing the validity of such
models. In particular, the determination of the difference of the electric
and magnetic dipole polarizabilities is of major importance. The value
of this difference found from the radiative $\pi^+$ meson photoproduction
from the proton at MAMI \cite{mami} is equal to
$11.6\pm 1.5_{stat}\pm 3.0_{syst}\pm 0.5_{mod}$ and close to the value
obtained from scattering of high energy $\pi^-$ mesons
off the Coulomb field of heavy nuclei in Serpukhov \cite{antip} assuming
$\alpha_{1\pi^-}=-\beta_{1\pi^-}$ and equal to $13.6\pm 2.8\pm 2.4$.
However, these values differ from the prediction of
chiral perturbation theory (ChPT) ($4.4\pm 1.0$ \cite{burgi}).
The experiment of the Lebedev Physical Institute on radiative pion
photoproduction from the proton \cite{lebed} has given
$\alpha_{1\pi^+}=20\pm 12$.
This value has large error bars but nevertheless shows a large discrepancy with regard
to the ChPT predictions as well.

Attempts to determine the charged pion dipole polarizabilities from 
the
reaction $\ggc$ suffered greatly from theoretical and experimental
uncertainties. The analyses \cite{bab,holst,kal} have been performed in the
region of low energy ($\sqrt{t}<500$~MeV, where $t$ is the square of the
total energy in $\g\g$ c.m. system). In this region the values of the 
experimental
cross sections of the process under consideration \cite{pluto,dm1,dm2,mark}
are very ambiguous. As a result, the values of $\alpha_{1\pi^{\pm}}$ found
lie in the interval 2.2--26.3. The analyses of the data of Mark II
\cite{mark} have given $\alpha_{1\pi^{\pm}}$ close to the ChPT result. However,
even changes of the dipole polarizabilities by more than 100\% are
compatible with the present error bars in the energy region considered
\cite{holst}.

The experimental information available so far for the dipole polarizabilities
of the charged pions is summarized in Table I.

\begin{table*}       
\caption{The experimental data presently available for the charged pion
dipole polarizabilities.}
\centering
\begin{tabular}{llcc} \hline\hline
\multicolumn{2}{l}{Experiments} & $\alpha_{1\pi^{\pm}}/\unit$ &
$\amc/\unit$  \\ \hline
\multicolumn{2}{l}{$\pi^{-}Z\rightarrow\g \pi^{-} Z$, Serpukhov (1983)
\cite{antip}} & $6.8\pm 1.4\pm 1.2$ &     \\  \hline
\multicolumn{2}{l}{$\gp$, Lebedev Phys.Inst. (1984) \cite{lebed}} &
$20\pm 12$ &     \\  \hline
\multicolumn{2}{l}{$\gp$, MAMI (2005) \cite{mami}} &   &
$11.6\pm 1.5_{stat}\pm 3.0_{syst}\pm 0.5_{mod}$ \\ \hline
\multicolumn{2}{l}{D. Babusci {\em et al.} (1992) \cite{bab}} &   &   \\
$\ggc$:   & PLUTO (1984) \cite{pluto} &
$19.1\pm 4.8\pm 5.7$ &      \\
    & DM 1 (1986) \cite{dm1} & $17.2\pm 4.6$ &     \\
    & DM 2 (1986) \cite{dm2} & $26.3\pm 7.4$ &     \\
    & Mark II (1990) \cite{mark} & $2.2\pm 1.6$ &    \\ \hline
\multicolumn{2}{l}{F. Donoghue, B. Holstein (1993) \cite{holst}} &   &   \\
$\ggc$:   & Mark II  & $2.7$     &   \\ \hline
\multicolumn{2}{l}{A. Kaloshin, V. Serebryakov (1994) \cite{kal}
} &   &   \\
$\ggc$:  & Mark II  &   & $5.25\pm 0.95$ \\ \hline\hline
\end{tabular}
\end{table*}

We have determined the charged pion polarizabilities from the analyses
of $\ggc$ reaction data in the energy region from threshold up to
2500 MeV.

An investigation of this process at middle energies was also carried out
in the framework of different theoretical models
\cite{oset,lee,penn,drech}. However,
they did not try to extract information about pion polarizabilities.

In Ref. \cite{fil1} we fitted the data \cite{mars} to the total cross
section of the process $\gg$, using dispersion relations (DRs) at fixed
$t$ with one subtraction for the invariant helicity amplitudes in the energy
region from 270 up to 2000 MeV. This analysis has allowed for the determination
of the $\pi^0$ dipole polarizabilities. In this work the $\s$ meson was
considered as an effective description of the strong $S$-wave $\pi\pi$
interaction using the broad Breit-Wigner resonance expression. The
parameters of such a $\s$ meson have been found from the fit to the
experimental data \cite{mars} in the energy region 270--825 MeV. As a
result, a good description of the experimental data was obtained for
$\sqrt{t}=270-1700$ MeV. However, this work predicted a strong rise
in the total cross section at higher energies in contradiction with
the experimental data.

In Ref. \cite{fil2} we showed that this discrepancy could be eliminated
in the energy region at least up to 2.25 GeV by regarding the quadrupole
$\pi^0$ meson polarizabilities as free parameters. With this aim, we have
constructed DRs at fixed $t$ with one subtraction, where the subtraction functions
were determined with the help of the DRs with two subtractions. The
subtraction constants were connected with the dipole and quadrupole
polarizabilities. The analysis of the experimental data \cite{mars,bien} for
the total cross section of the process $\gg$ was performed in the energy
region 270--2250 MeV. This analysis has resulted in the determination of
the quadrupole $\pi^0$ meson polarizabilities for the first time. In
addition, the values of the effective interaction radius of
the $f_2(1270)$ meson and its decay width into two
photons have been determined: $r_f=0.96\pm 0.01$ fm,
$\G_{f_2\to\g\g}=3.05\pm 0.11$ keV.

In the present work we construct the DRs similar to those of Ref. 
\cite{fil2} for the
amplitudes of the process $\ggc$. Using the DRs allows one to avoid
the problem of double counting and the subtractions in the DRs provide
a good convergence of the integrand expressions in these DRs and therefore
increases the reliability of the calculations.
These DRs are used to fit the experimental data
\cite{tpc,mark,cello,venus,aleph,belle}
in the energy region from threshold up to 2500 MeV.
As a result, we have found the dipole polarizabilities of the charged
pions and determined their quadrupole polarizabilities for the first time.
The value of the difference $\amc=13.0^{+2.6}_{-1.9}$ found is in good
agreement with the result of Refs. \cite{antip,mami} and with the 
predictions
of the dispersion sum rules (DSRs) \cite{fil2}. However this result is at
variance with the recent calculations in the framework of ChPT \cite{burgi}.

The paper is organized as follows. In Sec. II the DRs for the invariant helicity
amplitudes
of the process $\ggc$ are constructed. The determination of the charge pion
polarizabilities from the experimental data on the reaction $\ggc$ is
given in Sec. III. The discussion of the results obtained is presented in Sec.
IV. Conclusions are presented in Sec. V.

\section{Dispersion relations for the amplitudes of the process $\ggc$}

The process $\ggc$ is described by the following invariant
variables
\be
t=(k_1+k_2), \quad s=(p_1-k_1)^2, \quad u=(p_1-k_2)^2,
\ee
where $p_1(p_2)$ and $k_1(k_2)$ are the pion and photon
four-momenta.

We will consider the helicity amplitudes $\mpp$ and $\mm$ \cite{aber}.
These amplitudes have no kinematical singularities or zeros
and define the cross section of the process $\ggc$
as follows
\beq
&&\sigc=\frac1{128\pi^2}\sqrt{\frac{(t-4\m)}{t^3}}\left\{t^2 |\mpp|^2
\right. \nonumber \\
&& \left. +\frac{1}{16}t^2(t-4\m)^2\sin^4\tg|\mm|^2\right\},
\label{sig}
\eeq
where $\tg$ is the angle between the photon and the pion in the c.m. 
system of the process $\ggc$ and $\mu$ is the $\pi^{\pm}$ meson mass.

By constructing the DRs at fixed $t$ with one subtraction at $s=\m$
for the amplitude $\mpp$ we have:
\beq
&&Re \mpp (s,t)=Re \overline{M}_{++} (s=\m,t)+B_{++}(s,t) \nonumber \\
&&+\frac{(s-\m)}{\pi}\int\limits_{4\m}^{\infty}d\sp~Im\mpp(\sp,t)\left[
\frac{1}{(\sp-s)(\sp-\m)}\right. \nonumber \\
&&\left.-\frac{1}{(\sp-u)(\sp-\m+t)}\right],
\label{dr1}
\eeq
where $B_{++}$ is the Born term equal to
\be
B_{++}(s,t)=\frac{2e^2\m}{(s-\m)(u-\m)}
\ee
and
\be
\overline{M}_{++}(s=\m,t)=\mpp(s=\m,t)-B_{++}(s=\m,t).
\ee

Via crossing symmetry these DRs are identical to the DRs with two subtractions.

We determine the subtraction function $Re \overline{M}_{++}(s=\m,t)$
with the help of the DRs at fixed $s=\m$ with two subtractions using
crossing symmetry between the $s$ and $u$ channels
\beq
&&Re \ov{M}_{++}(s=\m,t)= \nonumber \\
&&=\ov{M}_{++} (s=\m,0)+\left.t\frac{d\ov{M}_{++} (s=\m,t)}{dt}
\right|_{t=0} \nonumber \\
&& +\frac{t^2}{\pi}\left\{P\int\limits_{4\m}^{\infty}
\frac{Im\mpp(\tp,s=\m)~d\tp}{\tp^2(\tp-t)}\right. \nonumber \\
&&\left. +\int\limits_{4\m}^{\infty}
\frac{Im\mpp(\sp,u=\m)~d\sp}{(\sp-\m)^2(\sp-\m+t)}\right\},
\label{sub}
\eeq
where $P$ denotes a principal value integral.

The subtraction constants $\ov{M}_{++}(s=\m,t=0)$ and
$\left.d\ov{M}_{++}(s=\m,t)/dt\right|_{t=0}$ are determined in terms of
differences of the dipole ($\amc$) and quadrupole ($\bmc$) polarizabilities
by taking into account the expressions of the sum and the difference of
the generalized electric and magnetic polarizabilities of any multipole
order through invariant amplitudes \cite{guias},
\beq
&&\ov{M}_{++}(s=\m,t=0)=2\pi\mu\amc, \nonumber \\
&&\left.\frac{d\ov{M}_{++}(s=\m,t)}{dt}\right|_{t=0}=\frac{\pi\mu}{6}\bmc.
\label{a-b}
\eeq

The DRs for the amplitude $\mm(s,t)$ have the same form as expressions
(\ref {dr1}) and (\ref{sub}) but
with the substitutions: $\mpp\to \mm$, $Im\mpp \to Im\mm$, and
$B_{++}\to B_{+-}=B_{++}/\m$.
The subtraction constants are equal in this case to
\beq
&&\ov{M}_{+-}(s=\m,t=0)=\frac{2\pi}{\mu}\apc, \nonumber \\
&&\left.\frac{d\ov{M}_{+-}(s=\m,t)}{dt}\right|_{t=0}=\frac{\pi}{6\mu}\bpc .
\label{a+b}
\eeq

To calculate a principal value integral in eq. (\ref{sub}) we use
the following expression:
\beq
&&t^2 P\int\limits_{4\m}^\Lambda
\frac{F(\tp)~d\tp}{\tp^2(\tp-t)}= t^2 \int\limits_{4\m}^\Lambda
\frac{(F(\tp)-F(t))~d\tp}{\tp^2(\tp-t)} \nonumber \\
&&+F(t)\left[\ln\frac{4\m (\Lambda-t)}{\Lambda (t-4\m)}-
t\frac{\Lambda-4\m}{4\m\Lambda}\right].
\eeq

\section{Determination of the charged pion polarizabilities}

The DRs for the charged pions are saturated  by the contributions of
the $\rho(770)$, $b_1(1235)$, $a_1(1260)$, and $a_2(1320)$ mesons in
the $s$ channel and $\s$, $f_0(980)$, $f_0(1370)$, $f_2(1270)$, and
$f_2(1525)$ in the $t$ channel.

The parameters of the $\rho$, $b_1$, $a_2$, $f_2(1270)$, and
$f_2(1525)$ mesons are given by the Particle Data Group \cite{pdg}.
The parameters of the
$f_0(980)$, $f_0(1370)$, and $a_1$ mesons are taken  as follows:

\noindent
$f_0(980)$: $m_{f_0}=980$ MeV \cite{pdg}, $\G_{f_0}=70$ MeV (the average
of the PDG \cite{pdg} estimate),
$\G_{f_0\to \g\g}=0.39\times 10^{-3}$ MeV \cite{pdg},
$\G_{f_0\to \pi\pi}=0.84\,\G_{f_0}$ \cite{anis};

\noindent
$f_0(1370)$: $m_{f_0(1370)}=1434$ MeV \cite{ait}, $\G_{f_0(1379)}=173$ MeV
\cite{ait},
$\G_{f_0(1370)\to \g\g}=0.54\times 10^{-5}$ MeV \cite{morg},
$\G_{f_0(1370)\to \pi\pi}=0.26\,\G_{f_0(1370)}$ \cite{bugg};

\noindent
$a_1(1260)$: $m_{a_1}=1230$ MeV \cite{pdg}, $\G_{a_1}=$425 MeV (the average
value of the PDG estimate \cite{pdg}),
$\G_{a_1\to \g\pi^{\pm}}=0.64$ MeV \cite{zel}.

The values of the effective radius of the
$f_2(1270)$ meson and its decay width into two photons are taken from our
analysis of the reaction $\gg$ \cite{fil2}: $r_f=0.96$~fm,
$\G_{f_2(1270)\to\g\g}=3.05$~keV.

For the $\s$ meson we use the values of the mass and the decay widths
found in Ref. \cite{fil1}: $m_{\s}=547$ MeV, $\G_{\s}=1204$ MeV,
$\G_{\s\to \g\g}=0.62$ keV.
It is worth noting that by regarding the decay widths $\G_{\s}$ and
$\G_{\s\to\g\g}$ as  free parameters in the present work,
we have obtained practically the same values for them as in Ref. \cite{fil1}.
This value of $\G_{\s\to \g\g}$ differs strongly from the result of
Ref. \cite{penn} ($\G_{\s\to \g\g}=(3.8\pm 1.5)$~keV). Using the latter
magnitude of the decay width in our fit led to
essential contradictions with the experimental data. On the other hand,
the value $\G_{\s\to \g\g}=0.62$ keV allowed Schumacher \cite{schum} to
obtain a reasonable value of the $\s$ meson contribution to the difference
of the nucleon dipole polarizabilities.

Expressions for the imaginary parts of the resonances under consideration
are given in Ref. \cite{fil2}.

The influence of the upper integration limit ($\Lambda$) in the DRs
on results of the calculations was investigated. They are not
changed for $\Lambda$ more than (5 GeV)$^2$.
In the present work we performed the integrations up to (12.5~GeV)$^2$.

The results of calculations are sensitive to a step length of integration
at $\tp <(4$~GeV$)^2$. We took the minimal integration step length in this
region which provided the stable result of the calculations.

The main contribution to the total cross section of the process $\ggc$
is given by the $t$-channel resonances and the pion polarizabilities.
For $s$-channel resonances, the largest contribution is given by the
$\rho$ meson. Its contribution is important in the energy region of
$\sqrt{t}>1$ GeV.

The best result of our fit to the experimental data for the total cross
section \cite{tpc,mark,cello,venus,aleph,belle} in the energy region
from threshold up to 2.5 GeV is presented in Fig. 1 by the solid
curve.
\begin{figure}
\epsfxsize=8.6cm   
\epsfysize=6.1cm     
\centerline{
\epsffile{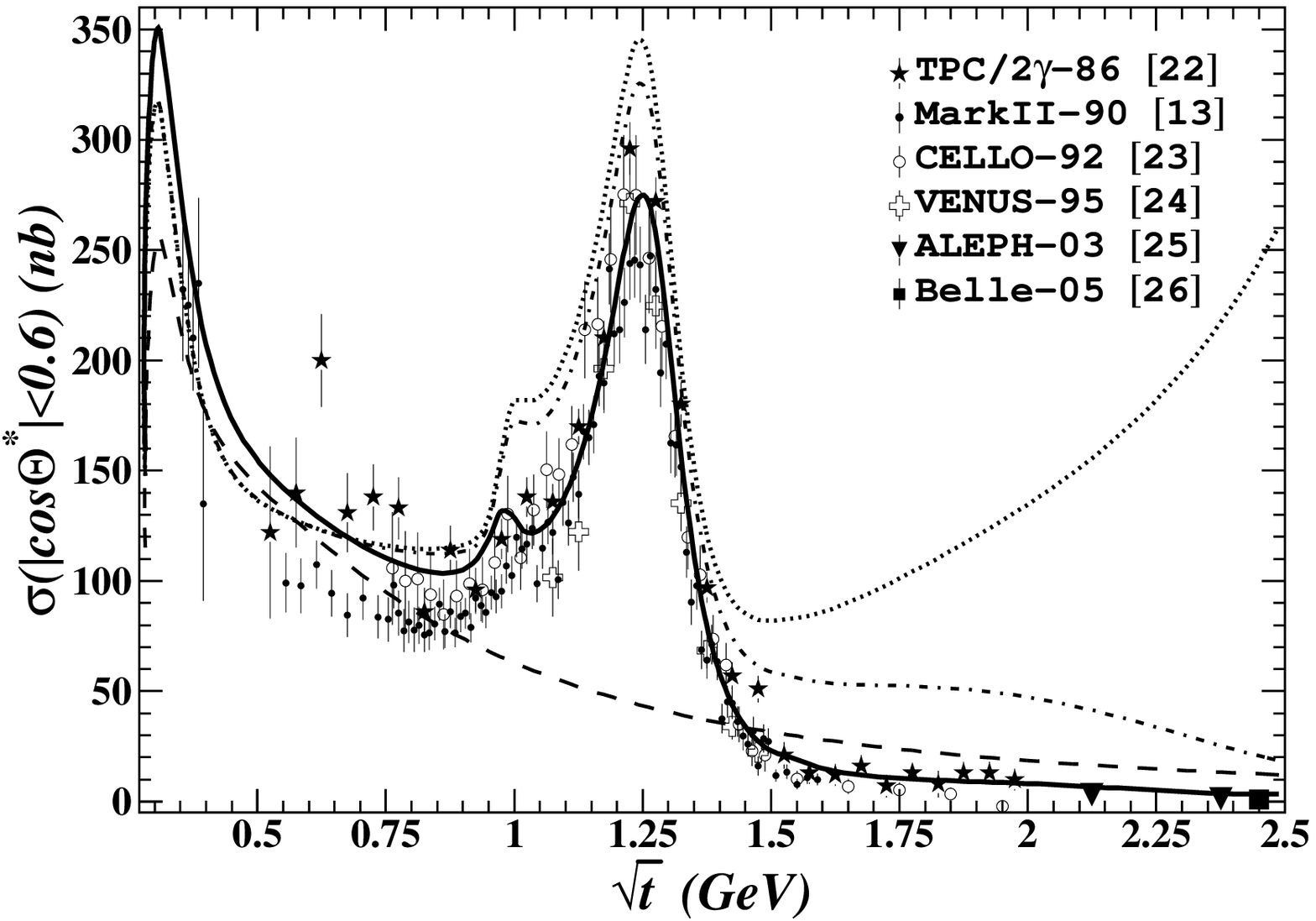}}
\caption{The total cross section for the reaction $\ggc$
(with $|\cos{\theta^*}|<0.6$). The solid curve is the best fit of the present
work. The dashed curve is the Born contribution. The dotted and
dashed-dotted curves are the results of a our calculation with
$\amc$ and $\apc$ from ChPT and the fit of the experimental data with
$\amc$ and $\apc$ fixed from ChPT, respectively.}
\end{figure}

We took into account all available data in this energy region which were
integrated over $|\cos{\theta^*}|<0.6$. The error bars in this figure are the
quadratic sum of statistical and systematic errors.

This solid curve well describes the experimental data in the whole energy region
under investigation. As a result of this fit, we have found the following
values for the charged pion polarizabilities:
\beq
&&\amc=13.0^{+2.6}_{-1.9}, \label{famc} \\
&&\apc=0.18^{+0.11}_{-0.02}, \label{fapc} \\
&&\bmc=25.0^{+0.8}_{-0.3}, \label{fbmc} \\
&&\bpc=0.133\pm 0.015. \label{fbpc}
\eeq

It should be noted that
the values of the sum of the dipole polarizabilities from Ref. \cite{ser}
are equal to $0.22\pm 0.06$ and $0.30\pm 0.04$ from the analysis of MarkII
and CELLO data, respectively.

The dashed curve in Fig. 1 is the Born term contribution.
The dotted curve is a result of calculations using the DRs
when $\bmc$ and $\bpc$ are equal to the values (\ref{fbmc}) and (\ref{fbpc})
but the dipole polarizabilities are taken from the ChPT calculations
\cite{burgi} as $\amc=4.4$ and $\apc=0.3$.
The dashed-dotted curve presents a result of the fit of the experimental data
when the quadrupole polarizabilities are the free parameters and the values
of the dipole  polarizabilities are fixed by the ChPT calculations
\cite{burgi}. This fit has given $\bmc=25.7$ and $\bpc=0.124$.
Both the last curves are close to the results of the calculations in Ref.
\cite{holst}
in the energy region up to 700 MeV; however, they differ strongly from all
experimental data on the total cross section at higher energies.

The fits of the data to the total cross section for the separate works
\cite{tpc,mark,cello,venus} are presented in Fig. 2. These
fits have given  lower and upper values of the charged pion polarizabilities
and were used to estimate
errors in Eqs. (\ref{famc})--(\ref{fbpc}) when the data of all works
\cite{tpc,mark,cello,venus,aleph,belle} were considered.
\begin{figure}
\epsfxsize=8.6cm
\epsfysize=7cm
\centerline{
\epsffile{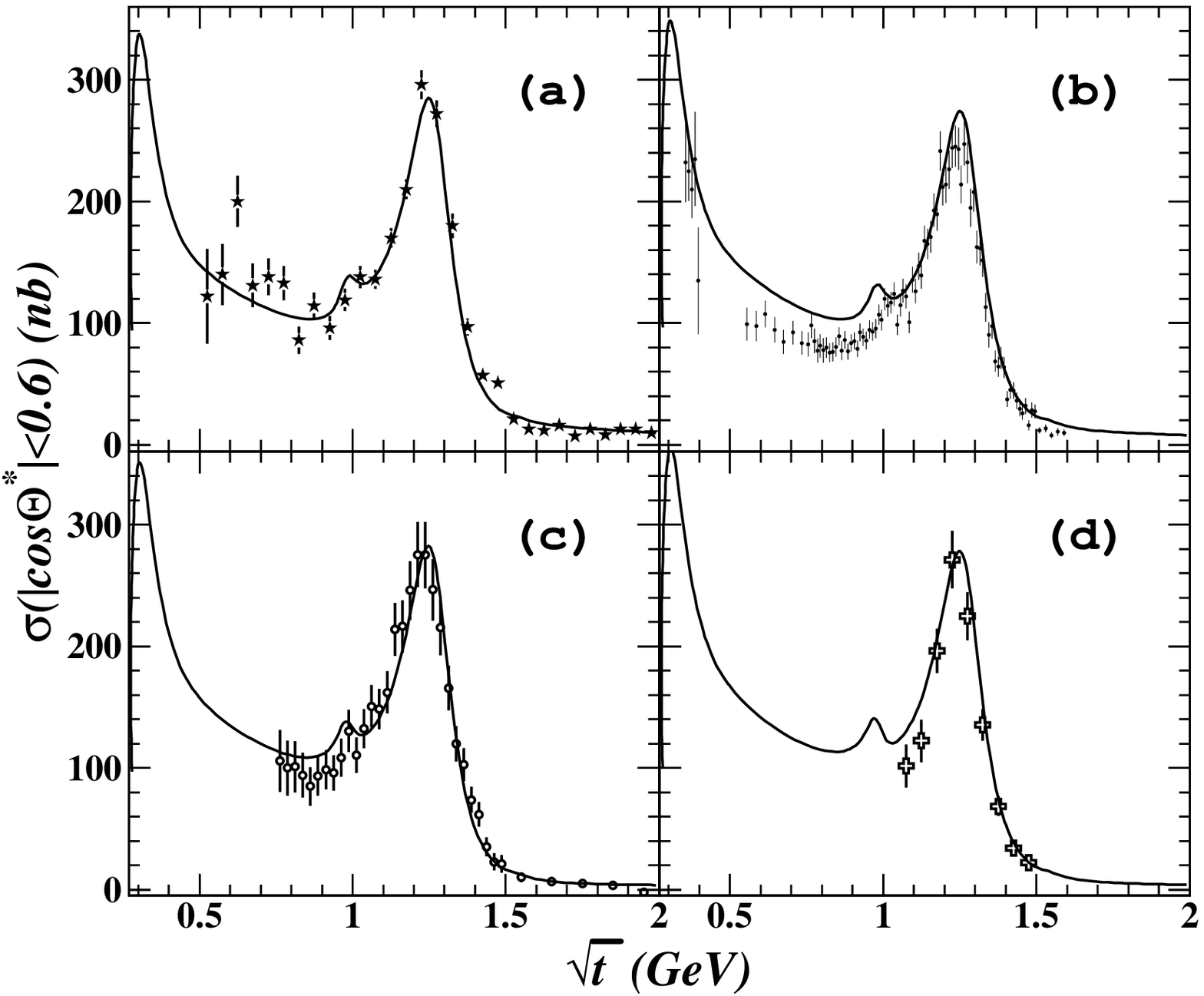}}
\caption{Fits to the total cross section of the individual experimental
works: (a) -- Ref. \cite{tpc}, (b) -- Ref. \cite{mark},
(c) -- Ref. \cite{cello}, (d) -- Ref. \cite{venus}.}
\end{figure}

The angular distributions of the differential cross section of the
process $\ggc$ at different energies are shown in Fig. 3.
\begin{figure}
\epsfxsize=8.6cm
\epsfysize=10.5cm
\centerline{
\epsffile{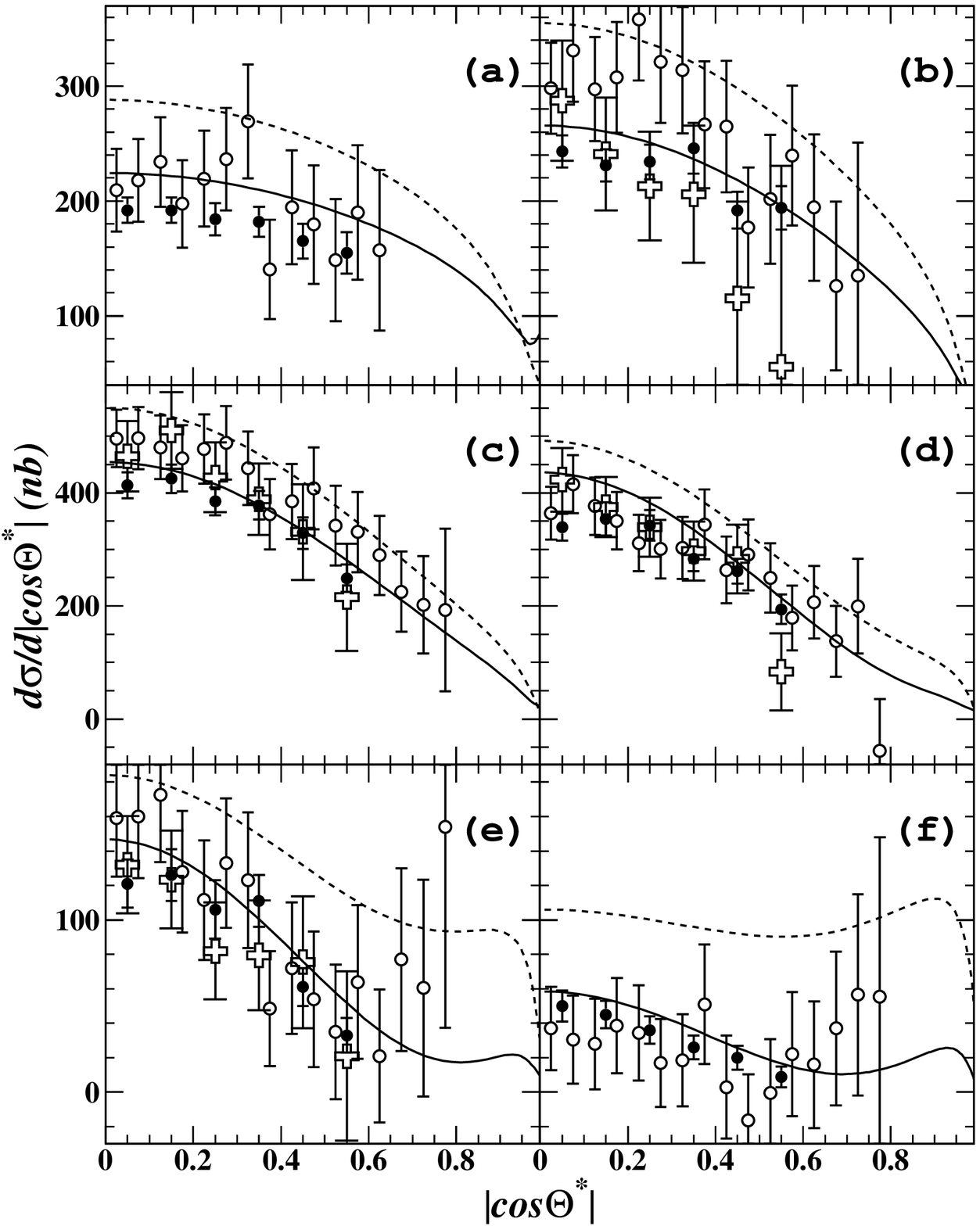}}
\caption{Angular distributions of the differential cross sections for the
following energy intervals: (a) -- 0.95--1.05 GeV, (b) -- 1.05--1.15 GeV,
(c) -- 1.15--1.25 GeV, (d) -- 1.25--1.35 GeV, (e) -- 1.35--1.45 GeV,
(f) -- 1.45--1.55 GeV.
The designations of the experimental data are the same
as in Fig. 1.}
\end{figure}
Presented in this figure are the experimental data of the MarkII, CELLO, and VENUS
collaborations. The differential cross sections of the CELLO
and VENUS collaborations were measured with 50-MeV energy intervals.
To compare these data with the MarkII ones we averaged them for 100-MeV
intervals.
The solid and dashed curves are the results of calculations
using our and the ChPT fits (the latter when the values of the dipole
polarizabilities are fixed by ChPT \cite{burgi})
to the total cross sections in Fig. 1, respectively.
This figure demonstrates a good description of the angular distributions by the
solid curves with the polarizability values (\ref{famc})--(\ref{fbpc}) found
in the present work.
However, the calculations with the dipole polarizabilities from
ChPT \cite{burgi} contradict these experimental data, particularly at
higher energies.

\section{Discussion}

The experimental results (\ref{famc})--(\ref{fbpc}) for the charged pion
dipole and quadrupole
polarizabilities and the predictions for these parameters from the
DSRs \cite{fil2} and the results of a two-loop analysis
(${\cal O}(p^6)$) \cite{burgi} for the dipole polarizabilities are listed
in Table II.
\begin{table}
\caption{The dipole and quadrupole polarizabilities of the charged pions
in units $\unit$ and $\unitq$, respectively.}
\centering
\begin{tabular}{cllr}\\ \hline\hline
         &present                &                &                  \\
         &analysis             &DSRs \cite{fil2} &ChPT \cite{burgi}\\ \hline
$\amc$   &$13.0^{+2.6}_{-1.9}$   &$13.60\pm 2.15$ &$4.4\pm 1.0$  \\ 
$\apc$   &$0.18^{+0.11}_{-0.02}$ &$0.166\pm 0.024$&$0.3\pm 0.1$  \\ 
$\bmc$   &$25.0^{+0.8}_{-0.3}$   &$25.75\pm 7.03$ &              \\ 
$\bpc$   &$0.133\pm 0.015$       &$0.121\pm 0.064$&         \\ \hline\hline
\end{tabular}
\end{table}
As seen from this Table, the values of the polarizabilities found are in a
good agreement with the DSR predictions \cite{fil2}.
The difference of the dipole polarizabilities of the charged pions
agrees very well with the results obtained from
the radiative photoproduction of $\pi^+$ from the proton at MAMI \cite{mami}
and from the scattering of high
energy $\pi^-$ mesons off the Coulomb field of heavy nuclei \cite{antip}.
However, this difference deviates
from the recent predictions of ChPT \cite{burgi}.

The approximate $SU(2)_L\times SU(2)_R\times U(1)_V$ chiral symmetry in
the two flavor sector of QCD results in a Ward identity, which relates
Compton scattering on a charged pion, $\g\pi^+\to\g\pi^+$, to radiative
charged-pion beta decay, $\pi^+\to e^+\nu_e\g$. The result obtained
for the charged pion dipole polarizabilities using ChPT at leading
nontrivial order (${\cal O}(p^4)$)
can be written in the form
\be
\alpha_{1\pi^\pm}=-\beta_{1\pi^\pm}=2\frac{e^2}{4\pi}\frac{1}{(4\pi F_\pi)^2
\mu}\frac{\overline{l}_6-\overline{l}_5}{6},
\ee
where $F_\pi=92.4$~MeV is the pion decay constant and
$(\overline{l}_6-\overline{l}_5)$
is a linear combination of scale-independent parameters of the Gasser and
Leutwyler Lagrangian \cite{gasser}. At the lowest nontrivial order
(${\cal O}(p^4)$) this difference is related to the ratio $\g=F_A/F_V$ of
the pion axial-vector form-factor $F_A$ and the vector form-factor $F_V$
of the radiative pion beta decay \cite{gasser}:
$\g=(\overline{l}_6-\overline{l}_5)/6$.
Using
$(\overline{l}_6-\overline{l}_5)=2.7\pm 0.4$ B\"urgi \cite{burgi} had
obtained at ${\cal O}(p^4)$
\be
\alpha_{1\pi^+}=2.7\pm 0.4.
\ee
This value is used in a two-loop analysis (${\cal O}(p^6)$) in Ref.
\cite{burgi}, the result is 
\beq
&& \apc=0.3\pm 0.1,  \label{p1} \\
&& \amc=4.4\pm 1.0.  \label{m1}
\eeq
It is worth noting that the degeneracy
$\alpha_{1\pi^+}=-\beta_{1\pi^+}$ has been removed
at ${\cal O}(p^6)$. The corresponding corrections amount to an 11\% (22\%)
change of the ${\cal O}(p^4)$ of the result for
$\alpha_{1\pi^+}$ $(\beta_{1\pi^+})$.
The effect of the new low-energy constants appearing at ${\cal O}(p^6)$
on the pion polarizabilities was estimated via resonance saturation by
including vector and axial-vector mesons. Their contribution was found
to be about 50\% of the two-loop result.

It should be noted that taking into account higher orders in the quark
mass expansion, Bijnens and Talavera have obtained
$(\overline{l}_6-\overline{l}_5=2.98\pm 0.33)$ \cite{bij} which would lead to
$\amc$ equal to 4.9 instead of 4.4, whereas the sum would remain the same as
in Eq. (\ref{p1}).
Nevertheless, the experimental results of
Ref. \cite{antip,mami} and the present ones differ from this
prediction of ChPT for $\amc$ and further theoretical
and experimental work is needed.

One of the reasons of such a deviation could be the neglect of the contribution
of the wide $\s$ meson in the ChPT calculations. As has been shown in
Ref. \cite{fil2}, this resonance gives a main contribution to DSRs for
$\amc$. This contribution essentially exceeds
the contribution of the vector and axial-vector mesons.

Moreover, the analysis of the recent PIBEPA result \cite{frlez} suggests 
an inadequacy of the present vector--axial-vector
$(V-A)$ description of the radiative beta decay, 
which would also reflect itself in an inadequacy of the ChPT description in 
its present form.

To compare with the situation for the $\pi^0$ mesons, we present here
Table III, where the experimental values of the $\pi^0$ meson dipole and
quadrupole polarizabilities are confronted with predictions of DSRs
\cite{fil1,fil2} and two-loop calculations in the frame of ChPT.
\begin{table}
\caption{The dipole and quadrupole polarizabilities of the $\pi^0$ meson
in units $\unit$ and $\unitq$, respectively. The ChPT data from
Ref. \cite{bell,ivan}.}
\centering
\begin{tabular}{clcc}\hline\hline
        &experiment       &DSRs \cite{fil2} &ChPT  \\ \hline
$\amn$   &$-1.6\pm 2.2$ \cite{fil1} &$-3.49\pm 2.13$ &$-1.9\pm 0.2$ \\
        &$-0.6\pm 1.8$ \cite{kal}  &                & \\ 
$\apn$   &$0.98\pm 0.03$ \cite{fil1}&$0.802\pm 0.035$&$1.1\pm 0.3$  \\
        &$1.00\pm 0.05$ \cite{ser} &                & \\ 
$\bmn$   &$39.70\pm 0.02$\cite{fil2}&$39.72\pm 8.01$ &$37.6\pm 3.3$  \\
$\bpn$   &$-0.181\pm 0.004$\cite{fil2}&$-0.171\pm 0.067$&0. 04 \\ \hline\hline
\end{tabular}
\end{table}

The experimental values of the
sum and the difference of the dipole polarizabilities of $\pi^0$ and the
difference of its quadrupole polarizabilities do not conflict within the
errors with the predictions  of DSRs and ChPT.
However, there are very big errors in the
experimental values for the difference of the dipole polarizabilities.
Therefore, it is difficult to draw a more unambiguous conclusion. For
the sum of the quadrupole polarizabilities of $\pi^0$, the DSR result
agrees well with the experimental value, but ChPT predicts a positive
value, in contrast to the experimental result. However, as was noted in
Ref. \cite{ivan}, this quantity was obtained in  a two-loop approximation,
which is a leading  order result for this sum, and one expects substantial
corrections to it from three-loop calculations.

\section{Conclusion}

The dispersion relations at fixed $t$ with one subtraction at $s=\m$ have been
constructed for the invariant helicity amplitudes of the reaction $\ggc$. The
subtraction functions were determined with the help of the DRs at fixed $s=\m$
with two subtractions at $t=0$, where the subtraction constants were
expressed through the dipole and quadrupole polarizabilities of the charged
pion. These DRs, where the sum and the difference of the dipole and the
quadrupole charged pion polarizabilities were free parameters, were used to fit
the experimental data to the total cross section of the process $\ggc$
in the energy region from threshold to 2.5 GeV.

As a result, the dipole and the quadrupole polarizabilities have been
found. The quadrupole charged pion polarizabilities have been determined
for the first time.
The value of the difference of the dipole polarizabilities
$\amc=13.0^{+2.6}_{-1.9}$ found agrees well with
results obtained at MAMI \cite{mami} and in Serpukhov \cite{antip} and with
the DSR predictions whereas it is at variance with the recent calculations 
in the frame work of ChPT \cite{burgi}.

The results of the calculations of the angular distributions for the process
under consideration, using the DRs constructed in the present work and the
values of the dipole and the quadrupole polarizabilities
(\ref{famc})--(\ref{fbpc}), are in a good agreement with the experimental data.

\begin{acknowledgments}

The authors would like to thank D.~Hornidge, A.~Thomas, and Th.~Walcher for
useful discussions. This research is part of the EU integrated 
infrastructure initiative hadronphysics project under contract number
RII3-CT-2004-506078 and was supported in part by 
the Russian Foundation for Basic Research (Grant No 05-02-04014).
\end{acknowledgments}



\begin{thebibliography}{99}
\bibitem{lvov} D. Babusci, G. Giordano, A.I. L'vov, G. Matone, and
A.M. Nathan, Phys. Rev. C {\bf 58}, 1013 (1998).
\bibitem{holst1} B.R. Holstein, D. Drechsel, B. Pasquini, and
M. Vanderhaeghen, Phys. Rev. C {\bf 61}, 034316 (2000).
\bibitem{mami} J. Ahrens {\em et al}.,
Eur. Phys. J. A {\bf 23}, 113 (2005); nucl-ex/0407011.
\bibitem{antip} Yu.M. Antipov {\em et al.}, Phys. Lett. {\bf B121}, 445 (1983).
\bibitem{burgi} U. B\"urgi, Nucl. Phys. B {\bf 479}, 392 (1997).
\bibitem{lebed} T.A. Aybergenov {\em et al.}, Czech. J. Phys. {\bf 36}, 948
(1986).
\bibitem{bab} D. Babusci {\em et al}., Phys. Lett. {\bf B277}, 158 (1992).
\bibitem{holst} J.F. Donoghue and B.R. Holstein, Phys. Rev. D {\bf 48}, 137
(1993).
\bibitem{kal} A.E. Kaloshin and V.V. Serebryakov, Z. Phys. C {\bf 64}, 689
(1994).
\bibitem{pluto} C. Berger {\em et al.}, (PLUTO Collaboration), Z. Phys. C
{\bf 26} 199 (1984).
\bibitem{dm1} A. Courau {\em et al.}, (DM1 Collaboration), Nucl. Phys. B
{\bf 271}, 1 (1986).
\bibitem{dm2} Z. Ajaltoni {\em et al.}, (DM2 Collaboration),
Phys. Lett. B {\bf 194}, 573 (1987).
\bibitem{mark} J. Boyer {\em et al.}, (Mark II Collaboration), Phys. Rev. D
{\bf 42}, 1350 (1990).
\bibitem{oset} J.A. Oller and E. Oset, Nucl. Phys. {\bf A629}, 739 (1998).
\bibitem{lee} C.-H. Lee, H. Yamagishi, and I. Zahed, Nucl. Phys. A {\bf 653},
185 (1999).
\bibitem{penn} M. Boglione, M.R. Pennington, Eur. Phys. J. C {\bf 9}, 11
(1999).
\bibitem{drech} D. Drechsel, M. Gorchtein, B. Pasquini, and M. Vanderhaeghen,
Phys. Rev. C {\bf 61}, 015204 (1999).
\bibitem{fil1} L.V. Fil'kov and V.L. Kashevarov, Eur. Phys. J. A {\bf 5}, 285
(1999).
\bibitem{mars} H. Marsiske {\em et al.}, (The Crystal Ball Collaboration),
Phys. Rev. D {\bf 41}, 3324 (1990).
\bibitem{fil2} L.V. Fil'kov and V.L. Kashevarov, Phys. Rev. C {\bf 72},
035211 (2005).
\bibitem{bien} J.K. Bienlein, Crystal Ball Contribution to the 9th 
International Workshop on Photon-Photon Collisions, San Diego, California, 
22--26 March 1992. Proceedings: Photon-Photon Collisions, edited by
D.O. Caldwell and H.P. Paar, River Edge, N.Y., World Scientific, 1992, 
p.241.
\bibitem{tpc} H. Aihara {\em et al.}, (TPC/$2\g$ Collaboration),
Phys. Rev. Lett. {\bf 57}, 404 (1986).
\bibitem{cello} H.J. Behrend {\em et al.}, (CELLO Collaboration), Z. Phys.
C {\bf 56}, 381 (1992).
\bibitem{venus} Fumiaki Yabuki {\em et al.}, (VENUS Collaboration),
J. Phys. Soc. Jap. {\bf 64}, 435 (1995).
\bibitem{aleph} A. Heister {\em et al.}, (ALEPH Collaboration), Phys. Lett.
B {\bf 569}, 140 (2003).
\bibitem{belle} H. Makazawa {\em et al.}, (Belle Collaboration), Phys. Lett.
B {\bf 615}, 39 (2005).
\bibitem{aber} H.A. Abarbanel and M.L. Goldberger, Phys. Rev. {\bf 165},
1594 (1968).
\bibitem{guias} I. Guiasu and E.E. Radescu, Ann. Phys. (N.Y.) {\bf 122}, 436
(1979).
\bibitem{pdg} S. Eidelman {\em et al.}, (PDG), Phys. Lett. B {\bf 592}, 1
(2004).
\bibitem{anis} V.V. Anisovich {\em et al.}, Phys. At. Nucl. {\bf 65},
1545 (2002).
\bibitem{ait} E.M. Aitala {\em et al.}, Phys. Rev. Lett. {\bf 86}, 765 (2001).
\bibitem{morg} D. Morgan and M.R. Pennington, Z. Phys. {\bf C48}, 623 (1990).
\bibitem{bugg} D.V. Bugg, A.V. Sarantsev, and B.S. Zou, Nucl. Phys.
{\bf B471}, 59 (1990).
\bibitem{zel} M. Zielinski {\em et al.}, Phys. Rev. Lett. {\bf 52}, 1195
(1984).
\bibitem{schum} M. Schumacher, Progr. Part. Nucl. Phys. {\bf 55}, 567 (2005).
\bibitem{ser} A.E. Kaloshin, V.M. Persikov, and V.V. Serebryakov,
Phys. Atom. Nucl. {\bf 57}, 2207 (1994).
\bibitem{gasser} J. Gasser and H. Leutwyler, Ann. Phys. (N.Y.) {\bf 158},
142 (1984).
\bibitem{bij} J. Bijnens and P. Talavera, Nucl. Phys. B {\bf 489}, 387 (1997).
\bibitem{frlez} E. Frle\u{z} {\em et al.}, Phys. Rev. Lett. {\bf 93},
181804 (2004).
\bibitem{bell} S. Bellucci, J. Gasser, and M.E. Sainio, Nucl. Phys.
B {\bf 423}, 80 (1994); B {\bf 431}, 413 (1994).
\bibitem{ivan} J. Gasser, M.A. Ivanov, and M.E. Sainio, Nucl. Phys.
B {\bf 728}, 31 (2005).

\end{thebibliography}
\end{document}